\documentclass[prl,twocolumn,superscriptaddress,showpacs]{revtex4}
\usepackage{graphicx,bm,amssymb,amsmath}

\newcommand{\Eq}[1]{Eq.~\eqref{#1}}
\newcommand{\eq}[1]{\eqref{#1}}
\newcommand{\Fig}[1]{Fig.~\ref{#1}}

\newcommand{\pdag}{{\phantom{\dagger}}}
\newcommand{\dx}{\partial_x}

\newcommand{\ket}[1]{\left\lvert{#1}\right\rangle}

\DeclareMathOperator{\sign}{sgn}
\DeclareMathOperator{\tsum}{\textstyle\sum}

\newcommand{\PRL}[3]{Phys. Rev. Lett.~\textbf{#1}, #2 (#3)}
\newcommand{\PRB}[3]{Phys. Rev. B~\textbf{#1}, #2 (#3)}
\newcommand{\PRA}[3]{Phys. Rev. A~\textbf{#1}, #2 (#3)}
\newcommand{\PR}[3]{Phys. Rev.~\textbf{#1}, #2 (#3)}
\newcommand{\Science}[3]{Science~\textbf{#1}, #2 (#3)}
\newcommand{\Nature}[3]{Nature~\textbf{#1}, #2 (#3)}
\newcommand{\JETP}[3]{Sov. Phys. JETP~\textbf{#1}, #2 (#3)}
\newcommand{\ZhETF}[3]{Zh. Eksp. Teor. Fiz.~\textbf{#1}, #2 (#3)}
\newcommand{\JMP}[3]{J. Math. Phys.~\textbf{#1}, #2 (#3)}

\newcommand{\JPB}[3]{J. Phys. B~\textbf{#1}, #2 (#3)}
\newcommand{\JSM}[2]{J. Stat. Mech.~#1 (#2)}

\newcommand{\etal}{\textit{et al.}}

\begin{document}

\title{Dynamics of  excitations in a one-dimensional Bose liquid }

\author{M. Khodas}
\affiliation{William I. Fine Theoretical Physics Institute and
School of Physics and Astronomy, University of Minnesota,
Minneapolis, MN 55455}
\author{M. Pustilnik}
\affiliation{School of Physics, Georgia Institute of Technology,
Atlanta, GA 30332}
\author{A. Kamenev}
\affiliation{ School of Physics and Astronomy, University of
Minnesota, Minneapolis, MN 55455}
\author{L.I. Glazman}
\affiliation{William I. Fine Theoretical Physics Institute and
School of Physics and Astronomy, University of Minnesota,
Minneapolis, MN 55455}

\begin{abstract}
We show that the dynamic structure factor of a one-dimensional Bose
liquid has a power-law singularity defining the main mode of collective 
excitations.  Using the Lieb-Liniger model, we evaluate the corresponding 
exponent as a function of the wave vector and the interaction strength.
\end{abstract}

\maketitle

Progress in the ability to manipulate ultracold atomic gases stimulates
the interest in fundamental properties of one-dimensional (1D) Bose
liquids~\cite{cold_gas_1D}. The quantity characterising the collective
excitations in these systems, the dynamic structure factor (DSF), is now
directly accessible experimentally using the Bragg spectroscopy
technique~\cite{Bragg}. Already the very first such 
measurements~\cite{Bragg_1D} clearly showed
that the resonance in DSF is wider in 1D than it is in higher dimensions.
The goal of this Letter is to elucidate the nature of the resonance in a
1D system of interacting bosons.

In the absence of interactions, bosons occupy the lowest-energy
single-particle state at zero temperature.  An external field that
couples to the particle density would excite bosons from the
ground state. The corresponding absorption spectrum reflects the
free boson's dispersion relation $\epsilon (q)$. Accordingly, 
DSF at zero temperature is given by
$S(q,\omega)\propto\delta\bigl(\omega-\epsilon(q)\bigr)$.

In dimensions higher than one, this behavior remains largely intact 
even in the presence of interactions. Bosons still form a condensate, 
and excitations of the system are very well described in terms of 
Bogoliubov quasiparticles~\cite{Pitaevskii}. Interactions merely affect 
their spectrum: 
$\epsilon(q) \propto q$ at small $q$.  
The quasiparticle decay rate scales with $q$ as 
$1/\tau_q\propto q^5$~\cite{Pitaevskii}; hence, the quasiparticle peak 
in $S(q,\omega)$ at $\omega = \epsilon(q)$ is well defined, 
$1/\tau_q\ll \epsilon(q)$. 

In 1D, the effect of interactions is dramatic: quantum fluctuations destroy 
the condensate.  Long-wavelength $(q\to 0)$ excitations of a 1D Bose liquid 
are often described in hydrodynamic approximation~\cite{Haldane} 
(see \cite{Cazalilla} for a recent review). However, the shape of the peak 
in DSF can not be addressed using this approach: in hydrodynamics the 
peak has zero width. 

In this Letter we study DSF of a 1D Bose liquid beyond the
hydrodynamic approximation. We consider the Lieb-Liniger (LL)
model~\cite{Lieb}: $N$ identical spinless bosons with contact
repulsive interaction placed on a ring with circumference $L$,
\begin{equation}
H = -  \frac{1}{2m}\sum_{i\,=1}^N \frac{\partial^2}{\partial
x_i^2} +\, c \sum_{ i<j} \delta(x_i - x_j)\, . \label{LL}
\end{equation}
The model is integrable~\cite{Lieb,Korepin}. The integrability allows 
one to relate the parameters of the hydrodynamic 
description~\cite{Haldane,Cazalilla}, the sound velocity $v$ and the 
parameter $K$, to the concentration $n=N/L$ and the dimensionless 
interaction strength $\gamma = mc/n$~\cite{K}. Finding dynamic 
correlation functions, such as DSF, in a closed form remains a 
challenge~\cite{Korepin}. The most impressive progress so far was
achieved by combining a finite-$N$ numerics with the Algebraic Bethe 
Ansatz~\cite{Caux}.  Here we study the singular behavior of DSF 
analytically. 

%%%%%%%%%%%%%%%%%%%%%%%%%%%%%%
\begin{figure}[h]
\includegraphics[width=0.9\columnwidth]{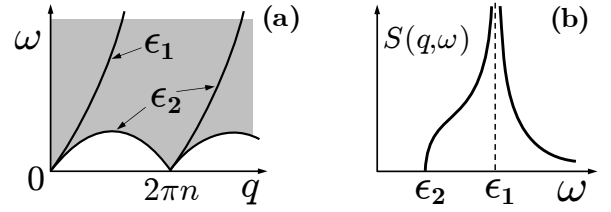}
\caption{ 
(a) Shaded area indicates the region in $(\omega,q)$-plane where 
$S(q,\omega)\neq 0$ at zero temperature. DSF exhibits power-law 
singularities along the solid lines.
(b) Sketch of the dependence of the structure factor on $\omega$
at a fixed $q<2\pi n$. 
}
\label{fig1}
\end{figure}
%%%%%%%%%%%%%%%%%%%%%%%%%%%%

DSF is defined by
\begin{equation}
S (q,\omega) = \int\!dx\,dt\,e^{i (\omega t-qx)}\,
\bigl\langle \rho (x, t) \rho (0, 0) \bigr\rangle,
\label{DSF}
\end{equation}
where $\rho(x) = \sum_i\delta(x-x_i)$ is the density operator. 
We show that DSF exhibits power-law singularities at the
Lieb's modes $\epsilon_{1,2}(q)$~\cite{Lieb,Lieb_modes}, see
\Fig{fig1}(a). In particular, DSF \textit{diverges} at
$\omega\to\epsilon_1(q)$ as
\begin{equation}
S(q, \omega) \sim \frac{m}{q} \,
\left|
\frac{\delta\epsilon}{\omega - \epsilon_1}
\right|^{\,\mu_1}\!
\bigl[ \theta(\epsilon_1-\omega) \,+ \,\nu_1
\theta(\omega-\epsilon_1) \bigr],
\label{LiebI}
\end{equation}
see \Fig{fig1}(b). Here
$\delta\epsilon(q)=\min\{\epsilon_1-\epsilon_2,vq\}$. Note that the
divergence occurs within the continuum. % spectrum of excitations.

The exponent $\mu_{1}$ and the coefficient $\nu_{1}$ in \Eq{LiebI}
depend on the dimensionless momentum $Q=q/mc$ and the interaction 
strength $\gamma$.  We were able to compute $\mu_1$ and $\nu_{1}$
in two limiting cases:
\begin{equation}
\mu_1 = 1 - (2K)^{-1},\quad \nu_1=1
\label{4}
\end{equation}
for $Q\gg (\gamma K)^{-1}\sim\max\{1,\gamma^{-1/2}\}$ 
and arbitrary $\gamma$, and
\begin{equation}
\mu_1=(\delta/\pi)\!\left(1- \delta/2\pi\right),
\quad
\nu_1=\frac{\sin(\delta^2\!/4\pi)}{\sin(\delta-\delta^2\!/4\pi)}
\label{5}
\end{equation}
for $\gamma\gg 1$ and arbitrary $Q$ (here $\delta=2\arctan Q$).
According to \Eq{5}, $\mu_{1}\approx 2Q/\pi$ at $Q\to 0$; we expect
that $\mu_{1}\propto Q$ at small $Q$ for any $\gamma$.

The line $\omega=\epsilon_1(q)$ has a ``replica'',
$\omega=\epsilon_2(q)$, at $q>2\pi n$. Here DSF does not
diverge, but still has a power-law non-analyticity. The singular
part of DSF has the form $\delta S(q,\omega)\propto
|\omega-\epsilon_2|^{\mu_2}$ with the exponent
\begin{equation}
\mu_2=2K+ (2K)^{-1} -1
 \label{mutwo1}
\end{equation}
at arbitrary $\gamma$ and $Q\gg\max\{1,\gamma^{-1}\}$, and
\begin{equation}
\mu_2=(\delta/\pi)(1+\delta/2\pi)
\label{mutwo2}
\end{equation}
at $\gamma\gg 1$ and arbitrary $Q$.

Lieb's hole-like (according to Bethe-ansatz classification) mode~\cite{Lieb_modes}
$\epsilon_2(q)$ serves as the lower boundary of the support of $S(q,\omega)$ 
at $q<2\pi n$, see \Fig{fig1}. Here DSF is given by
\begin{equation}
S(q, \omega) \sim \frac{m}{q} \, \left[ \frac{\omega -
\epsilon_2}{\delta\epsilon}\right]^{\mu_2}
\theta(\omega-\epsilon_2).
\label{LiebII}
\end{equation}
For $\gamma\gg 1$ the exponent $\mu_2$ here is given by  \Eq{mutwo2}.

Equations \eq{LiebI}-\eq{LiebII} represent the main result of this Letter.  
The shape of $S(q,\omega)$ near $\omega=\epsilon_1(q)$, see \Eq{LiebI}, 
differs qualitatively from the Lorentzian quasiparticle peak in higher dimensions.  
In 1D, the collective mode is characterized by a power-law divergence of 
$S(q,\omega)$. This divergence is protected by the integrability 
and associated with it absense of three-particle collisions~\cite{Korepin}. 
It is smeared only at a finite temperature $T$,
\begin{equation}
\max \bigl\lbrace S(q,\omega)\bigr\rbrace_{\text{fixed }q} \propto
T^{- \mu_1(q)},
\quad
T\ll\delta\epsilon.
\label{LiebI-finiteT}
\end{equation}
An apparent saturation of the height of the peak with the decrease of
$T$ would provide a direct measure of three-particle scattering
(absent in LL model) or recombination~\cite{recombination} rates.

In the remainder of the Letter we outline the derivation of the
above results. We start with the limit of large $q$.  Consider the state
$\rho^\dagger_q\ket{0}$, where $\rho^\dagger_q = \sum_k
\psi^\dagger_{k+q}\psi_k^\pdag$ is the Fourier component of the
density operator %in the second-quantized representation
($\psi^\dagger_p$ creates a boson with momentum $p$), and
$\ket{0}$ is the ground state of the Bose liquid. Without
interactions, all bosons in $\ket{0}$ occupy the single-particle state 
with $k=0$. The operator $\rho^\dagger_q$ annihilates one such
boson while creating another in the empty state
with momentum $q$. With interactions present, the occupation
number falls off~\cite{AG} rapidly with $k$ at $k\gtrsim mv$.
Therefore, for $q\gg mv$ the state $\rho^\dagger_q\ket{0}$ still
contains a single particle at momentum close to $q$, as well as a
``hole'' in the quasi-condensate with much smaller momentum. This
observation suggests to approximate
\begin{equation}
\rho^\dagger_q \approx \int\!dx\,d^\dagger\!(x)\psi(x),
\label{rho}
\end{equation}
where 
$d^{\,\dagger\!}(x) = L^{-1/2}\! \sum_{|k|<k_0}\!
e^{-ikx}\,\psi^\dagger_{q+k}$ 
creates a high-momentum particle and $\psi(x)$ creates a long-wavelength 
hole; here, $k_0\sim mv$ is the high-momentum cutoff. The $d$-particle 
is described by the Hamiltonian
\begin{equation}
H_d = \int\!dx\,d^\dagger (x)
\bigl[\epsilon_1(q) - i v_d \dx\bigr]
d(x),
\quad
v_d = q/m.
\label{8}
\end{equation}
Here we took into account that $\epsilon_1(p)\approx p^2\!/2m$ at
large $p$~\cite{Lieb_modes} and linearized the dispersion relation
around $p=q$.  We treat the long-wavelength bosons in the conventional
hydrodynamic approximation~\cite{Haldane,Cazalilla},
\begin{equation}
\psi(x) = \left[n + \pi^{-1}\dx\varphi\right]^{1/2} e^{i\vartheta(x)}.
\label{10}
\end{equation}
The fields $\varphi,\vartheta$ obey 
$[\varphi(x),\vartheta(y)] = i(\pi/2)\sign(x-y)$ and their dynamics is
governed by the Hamiltonian
\begin{equation}
H_0 = \frac{v_0}{2\pi}\int\!dx\!
\left[\frac{(\dx\varphi)^2}{K^2}
+ (\dx\vartheta)^2\right],
\quad
v_0 = \frac{\pi n}{m}\,.
\label{11}
\end{equation}
Eqs. \eq{DSF}, \eq{rho}, and \eq{10} yield DSF in the form
\begin{equation}
S(q,\omega) = \int\!dx\,dt\,e^{i\omega t}
\bigl\langle B(x,t)B^\dagger(0,0)\bigr\rangle
\label{12}
\end{equation}
with $B^\dagger(x)\propto d^\dagger(x) \,e^{i\vartheta(x)}$.
Evaluation of \Eq{12} with the quadratic Hamiltonian $H_d+H_0$ is
straightforward and yields \Eq{LiebI} with $\mu_1$ and $\nu_1$
given by \Eq{4}. The decomposition \Eq{rho} is applicable  
for $q\gg k_0\sim mv$, hence the restriction on $Q$ in \Eq{4}.
On the other hand, the constraint $|k|<k_0$ on the momentum of
$d$-particle limits the applicability of \Eq{LiebI}
to $|\omega-\epsilon_1(q)|\lesssim qv$.

We now extend the above derivation to the vicinity of the mode
$\epsilon_2(q)$ at $q\gg \max\{2\pi n, n\gamma\}$. At these
momenta, $\epsilon_2(q)=\epsilon_1(q-2\pi n)$ is a replica of mode
$\epsilon_1(q)$.  At a given energy $\omega\approx\epsilon_2(q)$
the relevant excitation includes, in addition to $d$-particle, the
$2\pi n$-momentum excitation of the quasi-condensate \Eq{11}.
In hydrodynamics~\cite{Haldane,Cazalilla}, such excitation
corresponds to $\psi(x)\propto e^{i\vartheta(x)-2i[\pi nx +
\varphi(x)]}$ instead of \Eq{10}. DSF is still given by
\Eq{12} with 
$B^\dagger(x)\propto d^\dagger(x)\,e^{i\vartheta(x)-2i[\pi nx +\varphi(x)]}$ 
and with the replacement $\epsilon_1\to\epsilon_2$ in \Eq{8}. Evaluation of
\Eq{12} then yields a power law for the singular part of
$S(q,\omega)$ with the exponent $\mu_2$ given by \Eq{mutwo1}.

At $\gamma\gg 1$, one can take an advantage of the exact mapping~\cite{CS}
of the LL model onto fermions with $\delta''(x)$ interaction. The mapping 
generalizes the famous duality~\cite{Tonks} between impenetrable bosons 
and free fermions and is based on the elementary identity
\begin{equation}
2\arctan\,(p/mc) = \pi - 2\arctan\, (mc/p).
\label{GO}
\end{equation}
The second term in the r.h.s. here is the scattering phase shift
$\theta_s(p)$ of the symmetric wave function~\cite{scattering} of
two particles with relative momentum $p$ interacting via
$V_B=c\,\delta(x)$ potential. Adding $\pi$ to $\theta_s$
converts the symmetric wave function into the antisymmetric one.
On the other hand, the l.h.s. of \eq{GO} is the phase shift
$\theta_a(p)$ of the antisymmetric wave function of two particles
interacting via potential $V_F=-2/(m^2c)\,\delta''(x)$. In view of
the integrability of the LL model, the two-particle phase shifts
contain a complete information about the Bethe ansatz wave
function of the many-body problem. Thus, for any \textit{bosonic}
eigenstate of the LL model \eq{LL} there is a dual
\textit{fermionic} eigenstate of the Hamiltonian
\begin{equation}
H_F = - \frac{1}{2m}\sum_{j=1}^N \frac{\partial^2}{\partial x_j^2}
- \frac{2}{m^2 c}\, \sum_{ i> j} \delta''(x_i - x_j)
\label{fermiHam}
\end{equation}
that has the same energy. The two wave functions coincide in one
of the sectors, say $x_1<x_2\ldots <x_N$, but differ by their
symmetry with respect to the permutation of particles'
coordinates. Since the density operator does not permute
particles, its matrix elements between any two many-body
eigenstates of \Eq{LL} are identical to those evaluated with the
corresponding dual eigenstates of $H_F$. In particular, the DSF
for the LL model coincides with that for the fermionic model
\Eq{fermiHam}.

It is convenient to rewrite \Eq{fermiHam} in the second-quantized
representation,
\begin{equation}
H_F= \sum_{p} \xi_p\, \psi^{\dagger}_p \psi_p
+ \sum_{k} \frac{V_k}{2L}\, \rho_k \rho_{-k}  \, ,
\quad
V_k= \frac{2k^2}{m^2 c}\,.
\label{secondquant}
\end{equation}
Here the operator $\psi^{\dagger}_p$ creates a fermion with
momentum $p$ and energy $\xi_p=p^2\!/2m$ and 
$\rho_k=\sum_p \psi^{\dagger}_{p-k}\psi_{p}$.

Strong repulsion between the original bosons corresponds to a weak
interaction in the dual fermionic model Eqs.~\eq{fermiHam},
\eq{secondquant}. In the limit $c\to\infty$ Eqs.~\eq{fermiHam},
\eq{secondquant} describe free fermions. In this limit the structure factor 
differs from zero only in a finite interval,
$\epsilon_2<\omega<\epsilon_1$ with $\epsilon_{1,2}(q) = v_0 q \pm
q^2\!/2m$. A weak ($\propto 1/\gamma$) residual  interaction between 
fermions leads to corrections to $\epsilon_{1,2}$; for example, the Fermi 
velocity $v_0$ is replaced by the sound velocity 
$v=v_0/K \approx v_0(1-4/\gamma)$~\cite{K,Lieb_modes}.  
Rather than discussing these modifications, we concentrate here on the 
singularities in $S(q,\omega)$.

DSF is proportional to the dissipative response to a field that couples 
to density. In the fermionic representation, the absorption of a quantum 
with energy $\omega$ and momentum $q$ is due to excitation of 
particle-hole pairs; there is just one such pair in the limit $c\to\infty$. 
At $\omega\to\epsilon_1$, the hole is created just below the Fermi level 
while the particle has momentum close to $k_F+q$; here $k_F=\pi n/m$ 
is the Fermi momentum. In the presence of interactions, such process 
is accompanied by a creation of multiple low-energy particle-hole pairs 
near the two Fermi points $p=\pm k_F$.  Similar to the well-known 
phenomenon of the Fermi edge singularity in the X-ray absorption spectra 
of metals~\cite{Mahan}, the proliferation of low-energy pairs leads to
power-law singularities in the response function at the edges of the 
spectral support.

The Fermi edge singularity relies crucially on the sharpness of the
distribution function which is smeared at a finite temperature, hence
\Eq{LiebI-finiteT}. Note that in an integrable model \Eq{fermiHam}
there is no relaxation of excited fermions~\cite{fermions} as 
three-particle collisions are absent.  Therefore, at $T=0$ there is no
smearing of the power-law singularities in $S(q,\omega)$ even at finite
$\omega$~\cite{fermions,fermions2}.

The derivation of Eqs.~\eq{LiebI}, \eq{5}, \eq{mutwo2}, and \eq{LiebII} 
follows the method of Ref.~\cite{fermions1}.  Consider first the limit
$\omega\to\epsilon_1$.  We truncate the continuum of single-particle 
states to three narrow subbands~\cite{fermions1} of the width $k_0\ll q$: 
$d$-subband around $p=q$ that hosts a single particle in the final state 
of the transition, and two subbands, $\alpha=\pm$, around the right/left 
Fermi points $p=\pm \,k_F$ that accommodate low-energy particle-hole 
pairs. After linearization of the spectrum within each subband, the resulting
effective Hamiltonian takes the form
\begin{equation}
H = H_d + H_0 + H_{int}
\label{16}
\end{equation}
with $H_d$ given by \Eq{8} and
\begin{equation}
H_0 = \int\!dx\,
\tsum_\alpha\psi_\alpha^\dagger(x)[-i\alpha v\dx]\,\psi_\alpha^\pdag(x)
\label{17}
\end{equation}
with $\psi_\alpha(x) = L^{-1/2}\!\sum_k e^{i(k-\alpha k_F)x}\,\psi_k$.
The last term in the r.h.s. of \Eq{16} describes interaction,
\begin{equation}
H_{int} = \sum_\alpha U_\alpha \!\!
\int\!dx\,\rho_\alpha(x)\rho_d(x),
\label{18}
\end{equation}
where $\rho_\alpha = \psi_\alpha^\dagger\psi_\alpha^\pdag$ and the
coupling constants $U_\alpha$ are related to the parameters of the
initial Hamiltonian~\eq{secondquant}, see \Eq{19} below.

Note that $H_{int}$ does not include the direct interaction
between the right and left movers.  Indeed, the corresponding
coupling constant $V_{2k_F} = 8\pi v_0/\gamma$ is small in the
limit $\gamma\gg 1$. In the absence of such interaction, the
remaining coupling constants $U_\alpha$ are set by the
requirement~\cite{fermions2} that the two-particle scattering
phase shifts for the effective Hamiltonian \eq{16}-\eq{18} with
linearized spectrum reproduce those for the original model
\eq{fermiHam}, \eq{secondquant}. In the latter case, the phase
shifts $\delta_\pm \equiv \theta_a(q+k_F\mp k_F)$
(see~\cite{scattering}) are given by
\[
\delta_\pm=  2\arctan\,
\frac{V_{q+k_F \mp k_F}}{2(v_d\mp v)}\, .
\]
In the limit $\gamma\gg 1$ taken at a constant $Q=q/mc$, one finds
$\delta_\pm=\delta$. In order to reproduce these phase shifts, the
coupling constants in \Eq{18} must be equal to
\begin{equation}
U_{\pm} = -  (v_d\mp v) \,\delta,
\quad
\delta = 2\arctan Q.
\label{19}
\end{equation}
In terms of the effective Hamiltonian \eq{16}-\eq{19}, the
structure factor is given by \Eq{12} with $B^\dagger(x) =
d^\dagger(x)\psi_+(x)$.  Following the steps familiar from the
theory of the Fermi-edge singularity~\cite{Schotte}, we
arrive at \Eq{LiebI} with the exponent given by \Eq{5}.

The power law \Eq{LiebII} with exponent of \Eq{mutwo2} is obtained
in a similar fashion. The only difference is that at $q<2k_F$ the
$d$-subband is centered at momentum $p=k_F-q$ (i.e., it is
below the Fermi level) and hosts a single \textit{hole} with velocity 
$v_d=v-q/m$. The hole is relatively slow, $|v_d|<v$, which 
leads~\cite{fermions1} to $S(q,\omega)=0$ at $\omega<\epsilon_2(q)$. 
With $q\to 2k_F-0$, the center of $d$-subband is approaching $-k_F$. 
At larger $q$, one returns to the particle-like $d$-subband, but 
the density operator $B^\dagger$ in \Eq{12} is now given by
$B^\dagger(x) = d^\dagger(x)\psi_-(x)$.

The ``bosonic'' route of evaluation of DSF described first,
and the ``fermionic'' one described second, have a common region 
of applicability corresponding to both $Q$ and $\gamma$ being large. 
In this limit the fermionic calculation yields $\mu_1\to 1/2$ and $\mu_2\to 3/2$, 
see Eqs.~\eq{5}, \eq{mutwo2}. This is in agreement with the strong-repulsion 
limit $(K\to 1)$ of the result of the bosonic calculation, see Eqs.~\eq{4}, \eq{mutwo1}.
Note that at the special point $q=2\pi n$ the exponent $\mu_2$ can be found 
using the hydrodynamic approximation (indeed, $\epsilon_2\to 0$ at $q\to2\pi
n$, hence $\omega\to\epsilon_2$ limit is accessible within the
effective low-energy description). The hydrodynamics yields
$\mu_2=K-1\approx 4/\gamma$~\cite{Astrakharchik,Cazalilla},
in agreement with the corresponding limit of \Eq{mutwo2}.

The peculiarity of 1D Bose liquid is that an arbitrarily weak
repulsion between particles destroys condensation. This renders the
perturbation theory developed for higher dimensions~\cite{Pitaevskii}
inapplicable. The well-known alternative method based on the
hydrodynamic description of low-energy excitations~\cite{Haldane} also
has its limitations, yielding infinitely narrow resonance in DSF
at small $q$~\cite{Cazalilla}. In this Letter we demonstrated the existence
of power-law singularities in DSF. The two complementary
methods of analytic evaluation of DSF developed here allowed to find the 
corresponding exponents in several regimes.
Evaluation of the exponents in the entire range of parameters remains a
challenging problem.

%%%%%%%%%%%%%%%%%%
\begin{acknowledgments}
We thank J.-S. Caux and D. Gangardt for numerous discussions.  
Research at the University of Minnesota is supported
by DOE (Grant DE-FG02-06ER46310) and by A.P. Sloan foundation.
Research at Georgia Tech is supported by NSF (Grant DMR-0604107).
\end{acknowledgments}


\begin{thebibliography}{99}
\bibitem{Lieb_modes} Lieb's particle-like (type I) mode
$\epsilon_1(q)$ resembles the Bogoliubov spectrum~\cite{Lieb}:
$\epsilon_1(q)\approx vq$ at small $q$ ($v$ is sound velocity) and
$\epsilon_1(q)\approx q^2\!/2m$ at large $q$.  For a weak
interaction $(\gamma\ll 1)$ the crossover between the two limits
occurs at $q\sim mv\sim n\sqrt{\gamma}$. The hole-like (type II) mode
$\epsilon_2(q)$ merges with $\epsilon_1(q)$ at small $q$.
\bibitem{scattering}
The symmetric and antisimmetric two-particle wavefunctions are defined as
$\Psi_s=\cos[(p|x|+\theta_s)/2]$ and $\Psi_a=\sign(x)\sin[(p|x|+\theta_a)/2]$,
respectively. Here $p$ and $x$ are relative momentum and coordinate, and
$\theta_{s,a}(p)$ are scattering phase shifts.
\end{thebibliography}
\end{document}